# Study of microwave resonances induced by bias lines of shunted Josephson junctions

U. Yilmaz, S. Razmkhah, R. Collot, J. Kunert, R. Stolz, P. Febvre

*Abstract*—Bias lines routed over a ground plane naturally form microstrip lines associated with the presence of a capacitance. This can lead to unwanted resonances when coupled to Josephson junctions. This work presents an electrical model of a shunted Josephson junction with its bias lines and pads, fabricated with the 1 kA/cm² RSFQ niobium process of the FLUXONICS Foundry [1][2]. A T-model is used to simulate the microwave behavior of the bias line, predict resonances and design resonance-free superconducting circuits. The I-V characteristics of three shunted Josephson junctions have been obtained from time-domain simulations done with JSIM [3] and show a good match with the experimentally observed resonance at 230 GHz up to the normal resistance branch, measured at 4.2 K. The influence of the position and value of a series resistor placed on bias lines is studied to damp unwanted resonances at the junction.

*Index Terms*— Josephson junction; RSFQ; Single Flux Quantum; superconducting electronics

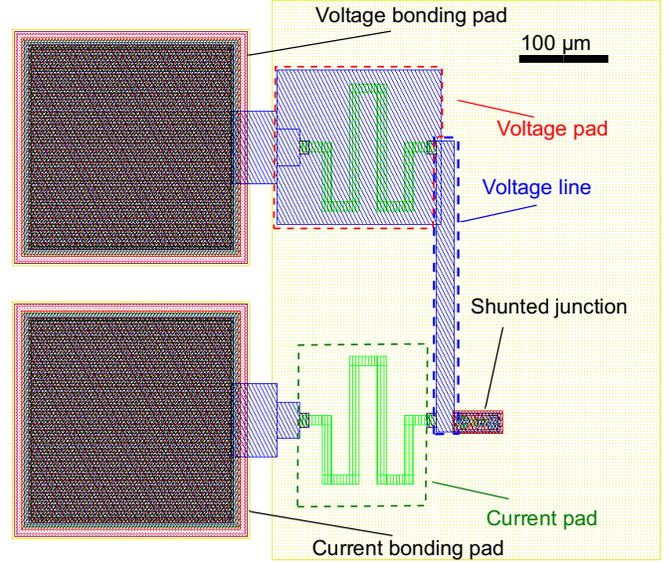

Fig. 1. Layout of a shunted Josephson junction fabricated with the 1 kA/cm² FLUXONICS Foundry process [1] connected to two bonding pads for 4-point measurements. Junction parameters : Ic=300 µA, Rshunt=1.47 Ω and Cshunt=1.5 pF. The ground plane is in yellow color. The bonding ground pads located at the chip's corners are not presented in this figure.

## I. INTRODUCTION

Application-specific superconducting circuits based on the Rapid Single Flux Quantum (RSFQ) digital technique are generally fabricated with a multilayered thin-film process which integrates one or more ground plane layers [1], [2], [4]. The ground plane, which often covers most of the chip surface, focuses the magnetic field under the superconducting wiring lines, which drastically lowers magnetic couplings between adjacent wires [5], [6]. Consequently bias lines and interconnects routed over the ground plane naturally form superconducting microstrip lines whose distributed nature can lead to strong resonances at some frequencies and modify the expected high frequency behavior of circuits, for instance for applications that require very high quality factors [7-11]. This happens whenever the microstrip lines present an inductive impedance that resonates with the intrinsic capacitance of the Josephson junction. In application such as magnetic field detection with SQUIDs [12], resonances are not always desirable and must thus be damped or removed [12-15]. They can also cause malfunctions in digital circuits for some operational conditions, for instance in presence of long Passive Transmission Lines (PTLs), or if high clock frequencies in the 100 GHz range are required, leading to circuit sizes close to the corresponding guided wavelength. This issue must be addressed since it will become more critical in the upcoming years when clock frequencies increase and PTLs are used more widely in the design of complex RSFQ circuits.

This work aims at predicting resonances and designing resonance-free superconducting circuits with design tools used in the RSFQ community, for investigation on resonances induced in RSFQ circuits. Simulation models of the Josephson junctions coupled to resonant cavities [16], [17] and physical models of the superconducting microstrip transmission lines [18], [19] have been studied in the past. These models are either not adapted to our devices or too complex to be implemented with our simulation tool. An electrical model of a shunted Josephson junction is presented with its bias lines and pads fabricated with the 1 kA/cm² RSFQ niobium process of the European FLUXONICS Foundry [1], [2]. Starting from the layout that includes the bias lines and pads shown in Fig. 1, the inductances and capacitances of superconducting layers are extracted with the InductEx commercial tool [20], and adjusted with simulations. The inductance and capacitance of the resistive layer is manually extracted. After introducing the

U. Yilmaz, S. Razmkhah, R. Collot and P. Febvre are with the IMEP-LAHC laboratory, Université Savoie Mont Blanc, 73376 Le Bourget du Lac, France (e-mail: ugur.yilmaz@univ-smb.fr, pascal.febvre@univ-smb.fr). J. Kunert and R. Stolz are with the Leibniz Institute of Photonic Technology, 07745 Jena, Germany.

fabrication process, a simple and compact model will be gained from these extractions and included in a SPICE netlist. Time-domain (TD) simulations of the netlist are done with the SPICE JSIM simulator [3] which comprises compact physical models of Josephson junctions and provides transient simulations of voltages, currents and phases at the desired nodes of the circuit under analysis. The measured *I-V* characteristics of the fabricated shunted junctions are compared with simulation results in a third step. Finally, a method to avoid resonances is presented.

## II. FABRICATION PROCESS

The reported test circuits have been fabricated by the FLUXONICS Foundry at Leibniz IPHT in Jena (Germany) using the standard process for RSFQ [1][2] (see Table I).
The layer stack consists of 13 layers, including three wiring layers of niobium (M0,M1,M2), one trilayer packet (TRI) in order to implement the Josephson junctions, two layers for resistors (R1,R2) and seven layers for isolation using either $SiO_2$ (I0B,I0C,I1B,I2,I3) or $Nb_2O_5$ (I0A,I1A). This process is well established for silicon wafers with a diameter of 100 mm and a thickness of 525 μm, which are covered with thermally oxidized silicon. The 600 nm thick $SiO_2$ layer acts as the first isolation.

An i-line stepper exposure tool with 5:1 reduction of structure size and maximum exposure field of 15 mm x 15 mm is used for photolithography of the first 12 layers. A 1:1 contact exposure for the whole wafer of the last layer (I3) serves for unambiguous identification of each chip. The first metal layer is a niobium ground plane M0 with a thickness of 200 nm. All niobium layers, the aluminum layer and the molybdenum layer were deposited by a DC magnetron plasma sputter tool. The niobium layers are reactive ion etched (RIE) with a $CF_4$ plasma, controlled by laser endpoint detection. The next layer I0A consists of $Nb_2O_5$ obtained by galvanic anodization, followed by I0B and I0C, each being a 125 nm-thick $SiO_2$ film deposited by a $SiH_4/O_2$ based PECVD process. The contact holes in the $SiO_2$ layers and the structures in R1 and R2 have been defined by lift-off technique. On top of this isolation layer the first wiring M1 is formed by a 250 nm thick niobium film. The next layer is the $Nb/Al/Al_2O_3/Nb$ (100 nm/10 nm/1-2 nm/30 nm) trilayer stack TRI. It defines the Josephson junction barrier. A controlled oxygen pressure and time for oxidation of aluminum surface defines the critical current density at 1 $kA/cm^2$. The trilayer is patterned in combination with RIE for niobium with physical plasma sputter etching for aluminum. A galvanic oxidation of the I1A layer defines the Josephson junction size for the desired critical current. The isolation is completed by layers I12 and I3, which are 150 nm-thick $SiO_2$ films. Between these $SiO_2$ layers an 80 nm-thick molybdenum layer R1 with a square resistance of 1 Ohm/square defines the shunt and bias resistors. A 350 nm-thick niobium layer M2 completes the process and allows for electrical interconnections of the layer stack. The whole structure is protected by a 400 nm-thick $SiO_2$ cover layer. On top of the bond pads a thermally evaporated 50 nm-thick gold layer R2 improves the conditions for wire bonding of the chips. This process and parameter set for the Josephson junctions is optimized for an operating temperature of 4.2 K.

TABLE I
LAYER DESCRIPTION OF THE FLUXONICS FOUNDRY RSFQ FABRICATION PROCESS USED FOR THE CHIP PRESENTED IN FIG. 1

| Layer name | Layer function | Thickness [nm] | Material |
| --- | --- | --- | --- |
| I3 | Protection | 200 | SiO |
| R2 | Bond pad | 50 | Au |
| M2 | Upper Wiring | 350 | Nb |
| I2 | Isolation | 150 | $SiO_2$ |
| R1 | Resistor | 80 | Mo |
| I1B | Isolation | 150 | $SiO_2$ |
| CUT | Remove additional M0 or M1 lines | - | - |
| I1A | Isolation | 70 | $Nb_2O_5$ |
| T1 | Josephson Junction | 60/12/30 | $Nb/Al$-$Al_2O_3/Nb$ |
| M1 | Lower Wiring | 250 | Nb |
| I0C | Isolation | 100 | $SiO_2$ |
| I0B | Isolation | 100 | $SiO_2$ |
| I0A | Isolation | 50 | $Nb_2O_5$ |
| M0 | Ground plane | 200 | Nb |

## III. MICROWAVE MODEL

An accurate model of a superconducting microstrip line that includes losses and kinetic inductance through the complex surface impedance requires specific simulation tools and theory [21]–[23]. At this stage it cannot be fully implemented in the JSIM simulator because of its dispersive properties. For the case of interest in this work, all resonant paths are cut into slices of approximately 25 μm, which are ten times shorter than the guided wavelength at the gap frequency (the gap voltage is about 2.5 mV), which is an upper limit for frequencies considered in this circuit. As shown in Fig. 2, the proposed model consists of three major blocks: a 175 μm-long voltage pad, a 320 μm-long voltage line, and a 525 μm-long bias

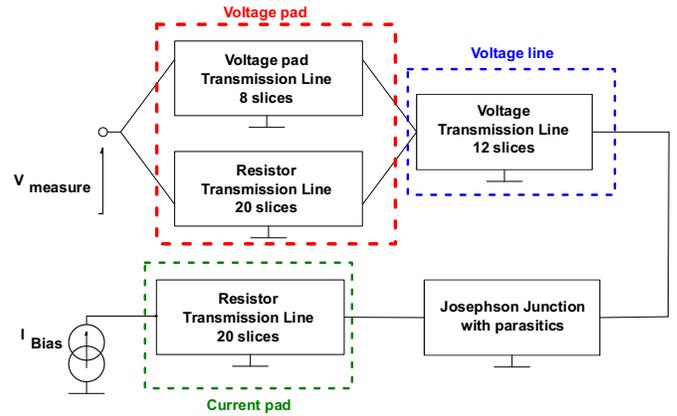

Fig. 2. Sketch of the model of the shunted Josephson connected to its voltage and current bias lines. A lumped-element model is used for current and voltage leads which are decomposed in a succession of lumped slices (see Fig. 3). Dotted color boxes match the ones of the layout presented in Fig. 1.

resistor used for the current pad and the voltage pad.

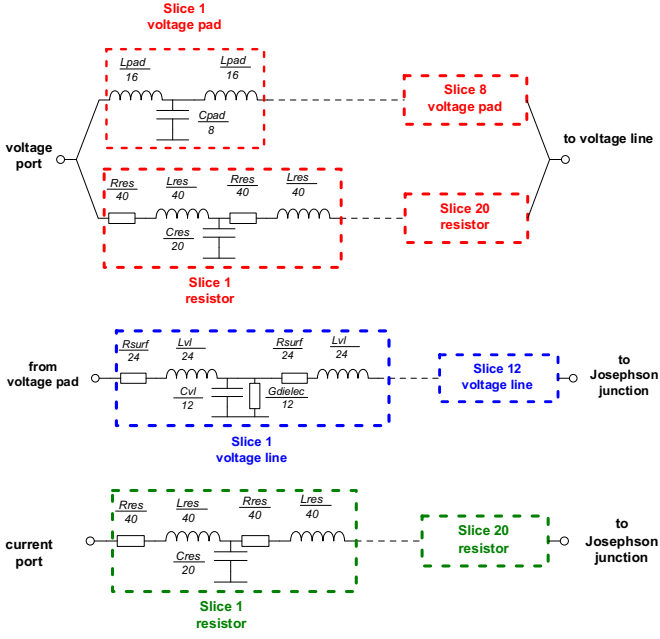

Fig. 3. Lumped T-model used to simulate the microstrip lines of the bias leads connected to the shunted Josephson junction. Dotted color boxes match the ones of the layout presented in Fig. 1. The lumped resistance, inductance and capacitance of the model are the ones of a 25µm-long slice of the corresponding box shown in Fig. 1.

As shown in Fig. 3, a lossy T-model is used for the voltage line. A lossless model is used for the voltage pad in parallel with a 52.5 Ω resistor sliced in 20 parts. The 52.5 Ω resistor is connected to the voltage pad only at its ends through a via between the R1 resistor layer and the M2 upper wiring layer (see Fig. 1).

This simple model neither considers the frequency-dependent surface impedance nor the frequency dependence of dielectric losses. Inductance and capacitance values are extracted with the InductEx commercial tool [8] for superconducting layers, and manually for the resistive layer, assuming a microstrip over a ground plane [24] as shown in equation (1).

$$L = \frac{120\pi l}{c}\left[\frac{1}{\frac{w}{h}+1.393+0.667\ln\left(\frac{w}{h}+1.44\right)}\right] \text{ for } \frac{w}{h}>1 \quad (1)$$

Therein, $L$ is the microstrip inductance, $c$ is the speed of light, $l$ is the microstrip length, $w$ is the microstrip width and $h$ is the height between the microstrip and the ground plane. The extracted inductance for the 525 µm-long, 10 µm-wide and 470 nm-thick dielectric layer is 26.6 pH, the capacitance is 0.45 pF.

IV. TIME DOMAIN SIMULATION

The measured *I-V* curve of the junction in Fig. 4 shows a critical current of 145 µA, 219 µA, and 226 µA for junctions with a critical current of 200 µA, 250 µA and 300 µA respectively. The difference between the design value and the measurement is due to the cool-down of the chip performed in a non-zero magnetic field. This measured critical current reduced by flux trapping has been used for JSIM simulations while the capacitance value is maintained for each junction at its nominal value, corresponding to the nominal specific capacitance of the 1 kA/cm² niobium FLUXONICS process. Time domain simulations with JSIM are used to reconstruct the *I-V* characteristics of the shunted Josephson junction. The bias current is maintained constant for 1 ns and the voltage response of the junction is averaged over the last 250 ps. Simulation parameters used for Fig. 4 are given in Table II. for the line parameters and Table III for the junctions including the parasitic effects caused by the shunt resistor.

TABLE II
PARAMETERS OF THE DIFFERENT TRANSMISSION LINE BLOCKS USED IN SIMULATIONS

| Parameters | Values used for simulation | Difference with extracted or designed values (%) |
|---|---|---|
| **Voltage pad** | | |
| Lpad | 14.1 pH | 0% |
| Cpad | 0.33 pF | 0% |
| **Voltage line** | | |
| Rsurf | 320 µΩ | 0% |
| Lvl | 14.1 pH | - 4% |
| Cvl | 0.33 pF | +10% |
| 1/Gdielec | 365 Ω | 0% |
| **Long resistor** | | |
| Rres | 52.5 Ω | 0% |
| Lres | 26.6 pH | 0% |
| Cres | 0.45 pF | 0% |

TABLE III
SIMULATION PARAMETERS FOR THE THREE JUNCTIONS OF FIG. 4

| Parameters | Values used for simulation | Difference from extracted or designed values (%) |
|---|---|---|
| **JJ 200 µA** | | |
| Critical current | 145 µA | -27.5% |
| Capacitance | 1.0 pF | 0% |
| Shunt resistor | 2.84 Ω | +29% |
| Parasitic inductance | 1.74 pH | 0% |
| **JJ 250 µA** | | |
| Critical current | 219 µA | -12.4% |
| Capacitance | 1.26 pF | 0% |
| Shunt resistor | 1.85 Ω | +3.9% |
| Parasitic inductance | 1.44 pH | 0% |
| **JJ 300 µA** | | |
| Critical current | 226 µA | -24.7% |
| Capacitance | 1.5 pF | 0% |
| Shunt resistor | 1.84 Ω | +25.2% |
| Parasitic inductance | 1.31 pH | 0% |

The simulated *I-V* curves present a good match at the resonance frequency of approximately 230 GHz if using parameters close to the designed ones. The stronger hysteresis that appears in the simulated *I-V* curves can be due to the junction model used in JSIM which only considers the Josephson current but neither takes into account the quasi-particle current nor the proximity effect. Measurement conditions in a closed-cycle GM cooler in presence of a periodically changing magnetic field, or flux trapping, can also be possible causes explaining the absence of hysteresis for a junction designed with a McCumber parameter of about 3.

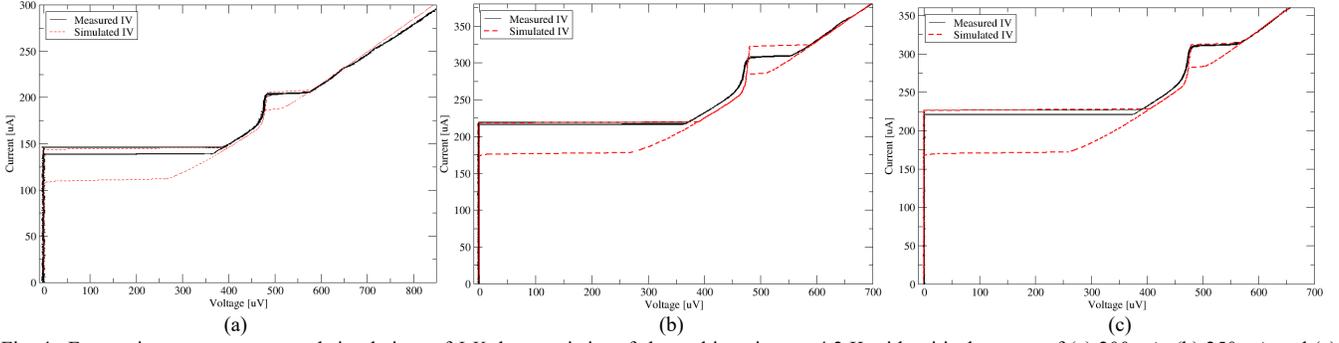

Fig. 4. Four-point measurements and simulations of *I-V* characteristics of shunted junctions at 4.2 K with critical current of (a) 200 µA, (b) 250 µA and (c) 300 µA.

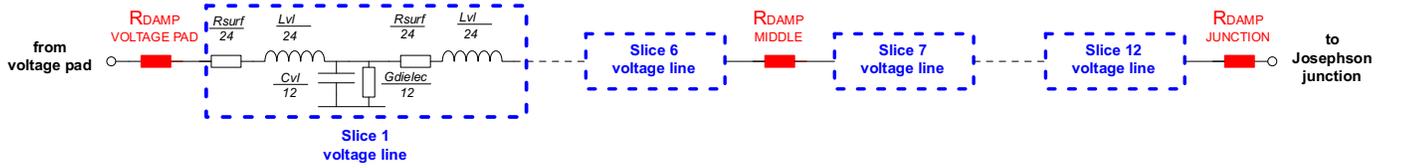

Fig. 5. Schematic of the damping resistance placement on the voltage line (see Fig. 1 and Fig. 2) used for simulations performed in Fig. 6. The damping resistor is placed successively at one of the three positions and simulations are performed accordingly.

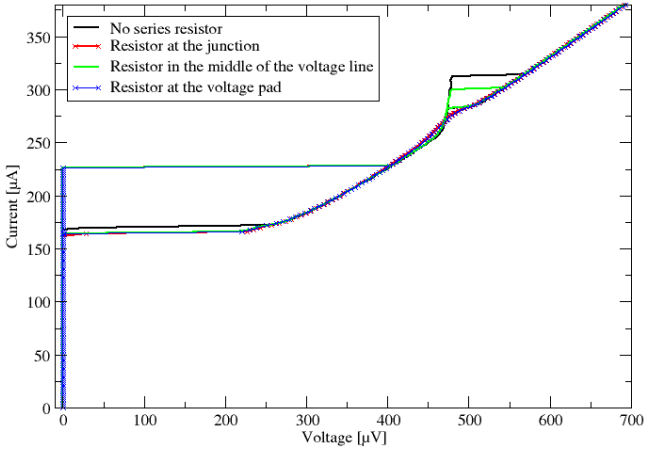

Fig. 6. Simulated *I-V* characteristics of the 300 µA shunted junction for a 1 Ω damping resistor placed at different positions along the voltage line.

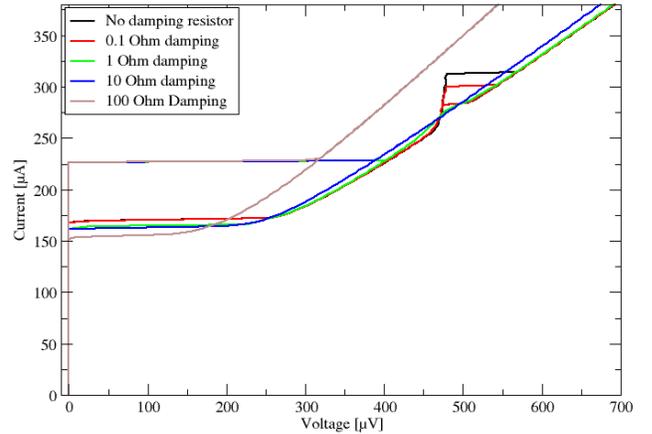

Fig. 7. Simulated *I-V* characteristics of the 300 µA shunted junction for different values of the damping resistor connected to the junction.

## V. Resonance-free design

In order to avoid unwanted resonances in superconducting integrated circuits one can tackle them on different ways: The parasitic capacitance can be suppressed by removing the ground plane at the expense of a higher inductance value. As inductance values are critical parameters for SFQ-based digital cells, removing the ground plane which focuses the magnetic flux and lowers the inductance is not always possible. The second option consists of placing a resistor in series along the bias line to damp the resonance.

In order to estimate the influence of the position and value of the damping resistor, simulations were performed for three positions of the resistor (Fig. 5) : at the voltage pad, in the middle of the voltage line, and directly connected to the junction, and for several values of the resistor (0.1 Ω, 1 Ω, 10 Ω and 100 Ω). From these multiple simulations, we can deduce that the most effective position to damp the resonances is by placing the series resistor as close as possible to the junction (Fig. 6). It should provide a resistance higher than or equal to the shunt resistor value of the Josephson junction (Fig. 7). Higher resistances of the series resistor give higher damping, but in this case the *IcRn* product observed in simulations (Fig. 7) is reduced due to the non-negligible contribution of the damping resistor.

## VI. Conclusion

The simple transmission line T-model with parameters extracted from the layout matches well the measured resonance at 230 GHz. The simple model can be easily implemented in the SFQ digital circuit design flow for a first order approximation of the resonance-induced perturbations. The current simple frequency-invariant model can still be improved by implementing a more accurate junction model and a frequency-dependent impedance of the superconducting microstrip lines. Damping the resonances induced by the microstrip lines can be done with a series resistor placed as close as possible to the Josephson junction, with values of a few ohms.

## Acknowledgment

This work has been partly funded by the French National Space Agency (CNES) and the Fond Européen De Développement Régional (FEDER) in the frame of the LSBB2020 project.